\documentclass[cup6b,epsf]{cupbook}

\usepackage{epsf}

\def\note #1]{{\bf #1]}}

\def\note #1]{{\bf #1]}}

\begin{document}

\author{Thomas Matheson\\
 Harvard-Smithsonian Center for Astrophysics \\
 60 Garden Street, Cambridge, MA  02138}

\chapter{The First Direct Supernova/GRB Connection:
 GRB~030329/SN~2003dh}

{\it Observations of gamma-ray burst (GRB) afterglows have yielded
   tantalizing hints that supernovae (SNe) and GRBs are related.  The
   case had been circumstantial, though, relying on irregularities in
   the light curve or the colors of the afterglow.  I will present
   observations of the optical afterglow of GRB 030329.  The early
   spectra show a power-law continuum, consistent with other GRB
   afterglows.  After approximately one week, broad peaks in the
   spectrum developed that were remarkably similar to those seen in
   the spectra of the peculiar Type Ic SN 1998bw.  This is the first
   direct, spectroscopic confirmation that at least some GRBs arise
   from SNe.
}

\section{Introduction}

The mechanism that produces gamma-ray bursts (GRBs) has been the
subject of considerable speculation during the four decades since
their discovery (see M{\'e}sz{\'a}ros 2002 for a recent review of the
theories of GRBs).  Optical afterglows (e.g.,~GRB~970228: Groot et
al.~1997; van Paradijs et al.~1997) opened a new window on the field
(see, e.g., van Paradijs, Kouveliotou, \& Wijers 2000).  Subsequent
studies of other bursts yielded the redshifts of several GRBs
(e.g.,~GRB~970508: Metzger et al.~1997), providing definitive evidence
for their cosmological origin.

Models that invoked supernovae (SNe) to explain GRBs were proposed
from the very beginning (e.g., Colgate 1968; Woosley 1993; Woosley \&
MacFadyen 1999).  A strong hint was provided by GRB~980425.  In this
case, no traditional GRB optical afterglow was seen, but a supernova,
SN~1998bw, was found in the error box of the GRB (Galama et
al. 1998a).  The SN was classified as a Type Ic (Patat \& Piemonte
1998), but it was unusual, with high expansion velocities (Patat et
al. 2001).  Other SNe with high expansion velocities (and usually
large luminosity as well) such as SN~1997ef and SN~2002ap are
sometimes referred to as ``hypernovae'' (see, e.g., Iwamoto et
al. 1998, 2000).
 
The redshift of a typical GRB is $z \approx 1$, implying that a
supernova component underlying an optical afterglow would be
difficult to detect.  At $z \approx 1$, even a bright core-collapse
event would peak at $R > 23$ mag.  Nevertheless, late-time deviations
from the power-law decline typically observed for optical afterglows
have been seen and these bumps in the light curves have been
interpreted as evidence for supernovae (for a recent summary, see
Bloom 2003).  Perhaps the best evidence that classical, long-duration
gamma-ray bursts are generated by core-collapse supernovae was
provided by GRB~011121.  It was at $z = 0.36$, so the supernova
component would have been relatively bright.  A bump in the light
curve was observed both from the ground and with \emph{HST} (Garnavich
et al.~2003a; Bloom et al.~2002).  The color changes in the light
curve of GRB~011121 were also consistent with a supernova (designated
SN~2001ke), but a spectrum obtained by Garnavich et al. (2003a) during
the time that the bump was apparent did not show any features that
could be definitively identified as originating from a supernova.  The
detection of a clear spectroscopic supernova signature was for the
first time reported for the GRB~030329 by Matheson et al. (2003a,
2003b), Garnavich et al. (2003b, 2003c), Chornock et al. (2003), and
Stanek et al.~(2003a).  Hjorth et al. (2003) also presented
spectroscopic data obtained with the VLT. In addition, Kawabata et
al. (2003) obtained a spectrum of SN~2003dh with the Subaru
telescope.

The extremely bright GRB~030329 was detected by instruments aboard
\emph{HETE II} at 11:37:14.67 (UT is used throughout this paper) on
2003 March~29 (Vanderspek et al.~2003).  Due to the brightness of the
afterglow, observations of the optical transient (OT) were extensive,
making it most likely the best-observed afterglow so far.  From the
moment the low redshift of 0.1685 for the GRB~030329 was announced
(Greiner et al.~2003), we started organizing a campaign of
spectroscopic and photometric follow-up of the afterglow and later the
possible associated supernova. Stanek et al.~(2003a) reported the
first results of this campaign, namely a clear spectroscopic detection
of a SN~1998bw-like supernova in the early spectra, designated
SN~2003dh (Garnavich et al. 2003c).  In this paper, I describe the
evidence for the supernova in the spectroscopy during the first two
months.  For a more complete discussion, see Matheson et al.~(2003c).

\begin{figure}
\begin{center}
\leavevmode\epsfxsize=10cm \epsfbox{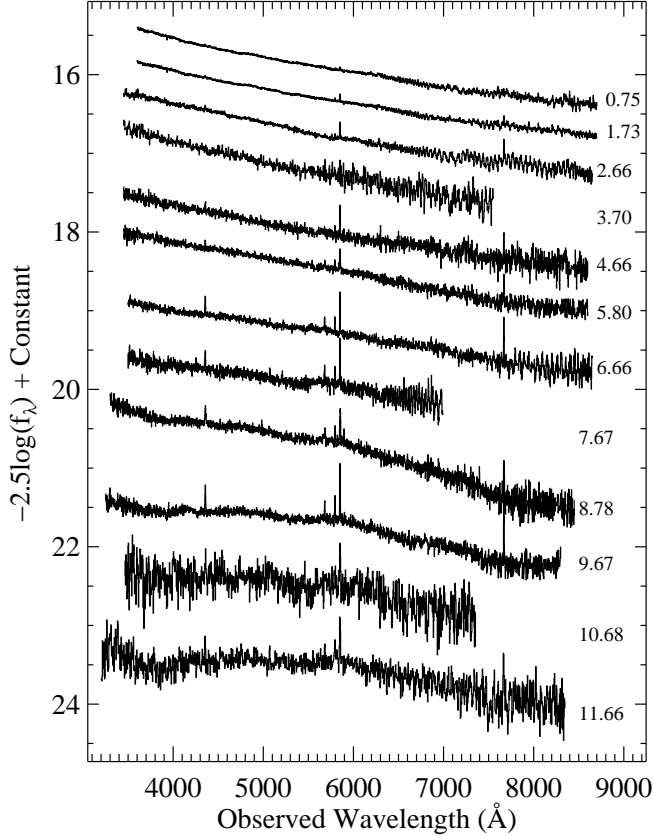}
\end{center}
\caption{Evolution of the GRB~030329/SN~2003dh spectrum, from
March~30.23 UT (0.75 days after the burst), to April 10.14 UT (11.66
days after the burst). The early spectra consist of a power-law
continuum with narrow emission lines originating from {\sc{H II}}
regions in the host galaxy at $z = 0.1685$. Spectra taken after
$\Delta T=6.66$ days show the development of broad peaks characteristic of
a supernova.
}
\label{fig:grb-all}
\end{figure}

\section{Spectra}

The brightness of the OT allowed us to observe the OT each of the 12
nights between March 30 and April 10 UT, mostly with the MMT 6.5-m,
but also with the Magellan 6.5-m, Lick Observatory 3-m, LCO du~Pont
2.5-m, and FLWO 1.5-m telescopes.  This provided a unique opportunity
to look for spectroscopic evolution over many nights.  The early
spectra of the OT of GRB~030329 (top of Figure \ref{fig:grb-all}) consist
of a power-law continuum typical of GRB afterglows, with narrow
emission features identifiable as H$\alpha$, [{\sc{O III}}]
$\lambda\lambda$4959, 5007, H$\beta$, and [{\sc{O II}}] $\lambda$3727
at $z = 0.1685$ (Greiner et al.~2003; Caldwell et al.~2003) probably
from {\sc{H II}} regions in the host galaxy.

Beginning at $\Delta T=7.67$ days, our spectra deviated from the pure
power-law continuum.  Broad peaks in flux, characteristic of a
supernova, appeared.  The broad bumps are seen at approximately
5000 \AA\ and 4200 \AA\ (rest frame). At that time, the spectrum of
GRB~030329 looked similar to that of the peculiar Type Ic SN~1998bw
a week before maximum light (Patat et al. 2001) superposed on a
typical afterglow continuum. Over the next few days the SN features
became more prominent as the afterglow faded and the SN brightened
toward maximum.

\begin{figure}
\begin{center}
\leavevmode\epsfxsize=10cm \epsfbox{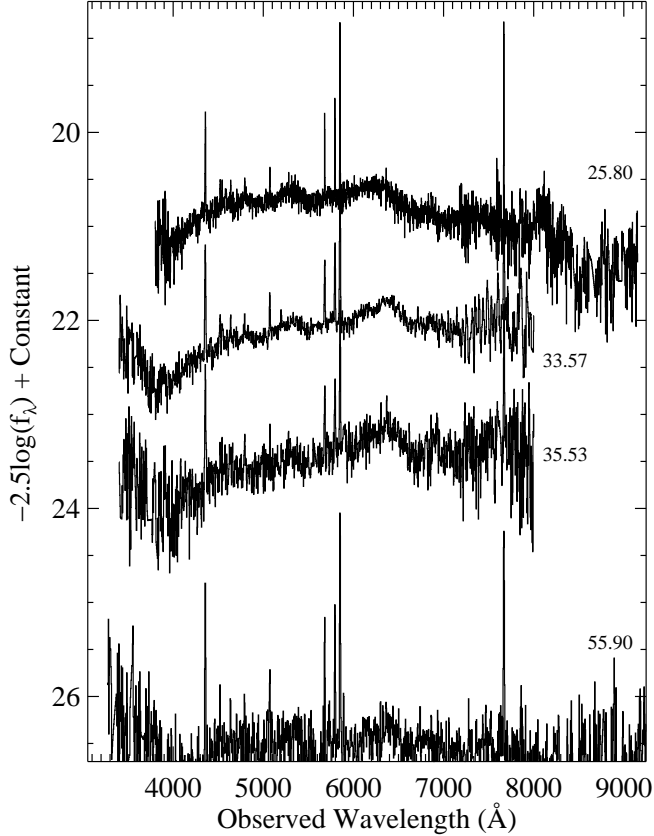}
\end{center}
\caption{Evolution of the GRB~030329/SN~2003dh spectrum, from April
24.28 UT (25.8 days after the burst), to May 24.38 (55.9 days after
the burst).  The power-law contribution decreases and the spectra
become more red as the SN component begins to dominate, although the
upturn at blue wavelengths may still be the power law.  The broad
features of a supernova are readily apparent, and the overall spectrum
continues to resemble that of SN~1998bw several days after maximum.
}
\label{fig:spectra-later}
\end{figure}

Later spectra obtained on April 24.28, May 2.05, May 4.01, and May
24.38 continue to show the characteristics of a supernova.  As
the power-law continuum of the GRB afterglow fades, the supernova
spectrum rises, becoming the dominant component of the overall
spectrum (Figure \ref{fig:spectra-later}).

\section{Separating the GRB from the Supernova}

To explore the nature of the supernova underlying the OT, we modeled
the spectrum as the sum of a power-law continuum and a peculiar Type
Ic SN.  Specifically, we chose for comparison SN~1998bw (Patat et
al. 2001), SN~1997ef (Iwamoto et al. 2000), and SN~2002ap (using our
own as yet unpublished spectra, but see, e.g., Kinugasa et al. 2002;
Foley et al. 2003).  We had 62 spectra of these three SNe, spanning
the epochs of seven days before maximum to several weeks past.  For
the power-law continuum, we chose to use one of our early spectra to
represent the afterglow of the GRB.  The spectrum at time $\Delta
T=5.80$ days was of high signal-to-noise ratio (S/N), and suffers from
little fringing at the red end.  Therefore, we smoothed this spectrum
to provide the fiducial power-law continuum of the OT for our model.

\begin{figure}

\begin{center}
\leavevmode\epsfxsize=10cm \epsfbox{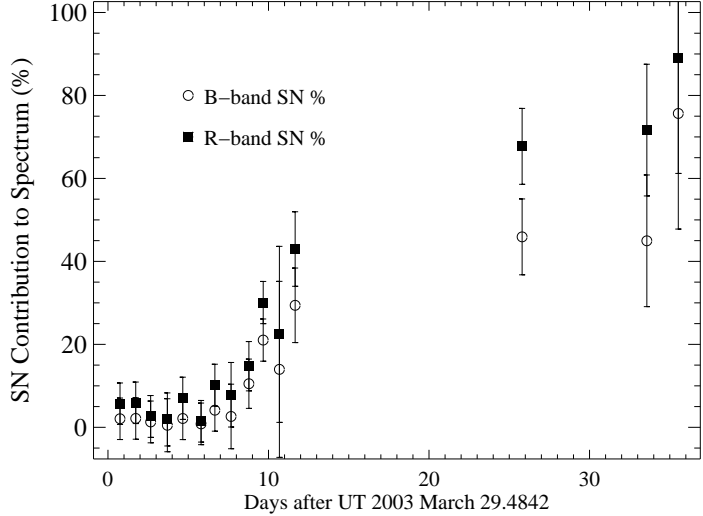}
\end{center}
\caption{Relative contribution of a supernova spectrum to the
GRB~030329/SN~2003dh afterglow as a function of time in the $B$
(\emph{open circles}) and $R$ (\emph{filled squares}) bands.  Using
the technique described in the text, we derive a best fit to the
afterglow spectrum at each epoch with the fiducial power-law continuum
and the closest match from our set of peculiar SNe Ic.  We then
synthesize the relative $B$-band and $R$-band contributions.  There is
some scatter for the early epochs due to noise in the spectra, but a
clear deviation is evident starting at $\Delta T=7.67$ days, with a
subsequent rapid increase in the fraction of the overall spectrum
contributed by the SN.  Errors are estimated from the scatter when the
SN component is close to zero ($\Delta T < 6$ days) and from the scale
of the error in the least-squares minimization.
}

\label{fig:snfrac}
\end{figure}

To find the best match with a supernova spectrum, we compared each
spectrum of the afterglow with the sum of the fiducial continuum and a
spectrum of one of the SNe in the sample.  We performed a
least-squares fit, allowing the fraction of continuum and SN to vary,
finding the best combination of continuum and SN for each of the SN
spectra.  The minimum least-squares deviation within this set was then
taken as the best SN match for that epoch of OT observation.  Figure
\ref{fig:snfrac} shows the relative contribution to the OT spectrum by
the underlying SN in the $B$ and $R$ bands as a function of $\Delta
T$.

Within the uncertainties of our fit, the SN fraction is consistent
with zero for the first few days after the GRB.  At $\Delta T=7.67$
days, the SN begins to appear in the spectrum, without strong evidence
for a supernova component before this.  When the fit indicates the
presence of a supernova, the best match is almost always SN~1998bw.
The only exceptions to this are from nights when the spectrum of the
OT are extremely noisy, implying that less weight should be given to
those results.  The least-squares deviation for the spectra that do
not match SN~1998bw is also much larger.

Our best spectrum (i.e., with the highest S/N) from this time when the
SN features begin to appear is at $\Delta T=9.67$ days.  For that
epoch, our best fit is 74\% continuum and 26\% SN~1998bw (at day $-6$
relative to SN $B$-band maximum).  The next-best fit is SN~1998bw at
day $-$7.  Using a different early epoch to define the reference
continuum does not alter these results significantly.  It causes
slight changes in the relative percentages, but the same SN spectrum
still produces the best fit, albeit with a larger least-squares
deviation.

\begin{figure}
\begin{center}
\leavevmode\epsfxsize=10cm \epsfbox{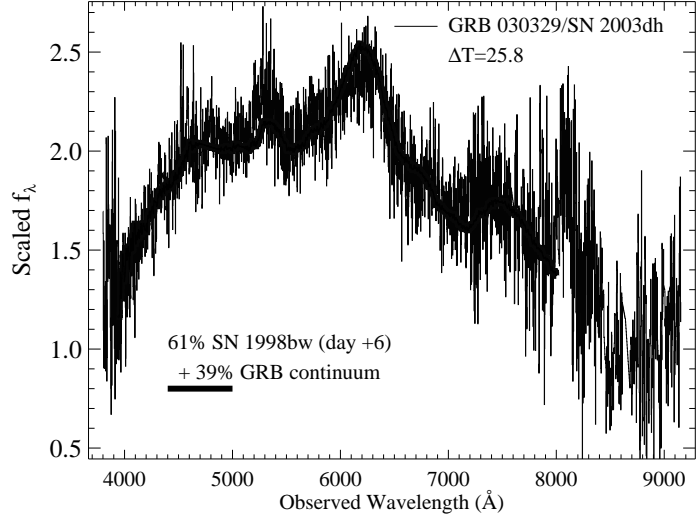}
\end{center}
\caption{Observed spectrum (\emph{thin line}) of the
GRB~030329/SN~2003dh afterglow at $\Delta T=25.8$ days.  The model spectrum
(\emph{thick line}) consists of 39\% continuum and 61\% SN~1998bw from
6 days after maximum.
}
\label{fig:day25}
\end{figure}

The SN fraction contributing to the total spectrum increases steadily
with time.  By $\Delta T=25.8$ days, the SN fraction is $\sim$ 61\%,
with the best-fit SN being SN~1998bw at day +6 (Figure \ref{fig:day25}).
The SN percentage at $\Delta T=33.6$ days is still about 63\%, but the
best match is now SN~1998bw at day +13.  The rest-frame time
difference between $\Delta T=9.67$ days and $\Delta T=25.8$ days is
13.8 days ($z = 0.1685$).  For the best-fit SN spectra from those
epochs, SN~1998bw at day $-$6 and SN~1998bw at day +6 respectively,
the time difference is 12 days.  The rest-frame time difference
between $\Delta T=25.8$ days and $\Delta T=33.6$ days is 6.7 days,
with a time difference between the best-fit spectra for those epochs
of 7 days.  The spectral evolution determined from these fits
indicates that SN~2003dh follows SN~1998bw closely, and it is not as
similar to SN~1997ef or SN~2002ap.  The analysis by Kawabata et
al. (2003) of their May 10 spectrum gives a phase for the spectrum of
SN~2003dh that is consistent with our dates, although they do consider
SN~1997ef as a viable alternative to SN~1998bw as a match for the SN
component in the afterglow.

The spectra of SN~1998bw (and other highly energetic SNe) are not
simple to interpret.  The high expansion velocities result in many
overlapping lines so that identification of specific line features is
problematic for the early phases of spectral evolution (see, e.g.,
Iwamoto et al. 1998; Stathakis et al. 2000; Nakamura et al. 2001;
Patat et al. 2001).  This includes spectra up to two weeks after
maximum, approximately the same epochs covered by our spectra of
SN~2003dh.  In fact, as Iwamoto et al. (1998) showed, the spectra at
these phases do not show line features.  The peaks in the spectra are
due to gaps in opacity, not individual spectral lines.  Detailed
modeling of the spectra can reveal some aspects of the composition of
the ejecta (Mazzali et al. 2003).

If the $\Delta T=9.67$ days spectrum for the afterglow does match
SN~1998bw at day $-$6, then limits can be placed on the timing of the
supernova explosion relative to the GRB.  The rest-frame time for
$\Delta T=9.67$ days is 8.2 days, implying that the time of the GRB
would correspond to $\sim$14 days before maximum for the SN.  The rise
times of SNe Ic are not well determined, especially for the small subset
of peculiar ones.  Stritzinger et al. (2002) found the rise time of
the Type Ib/c SN~1999ex was $\sim$18 days (in the $B$ band), while
Richmond et al. (1996) reported a rise time of $\sim$12 days (in the
$V$ band) for the Type Ic SN~1994I.  A rise time of $\sim$14 days for
SN~2003dh is certainly a reasonable number.  It also makes it
extremely unlikely that the SN exploded significantly earlier or later
than the time of the GRB, most likely within $\pm 2$ days of the GRB
itself.

\begin{figure}
\begin{center}
\leavevmode\epsfxsize=10cm \epsfbox{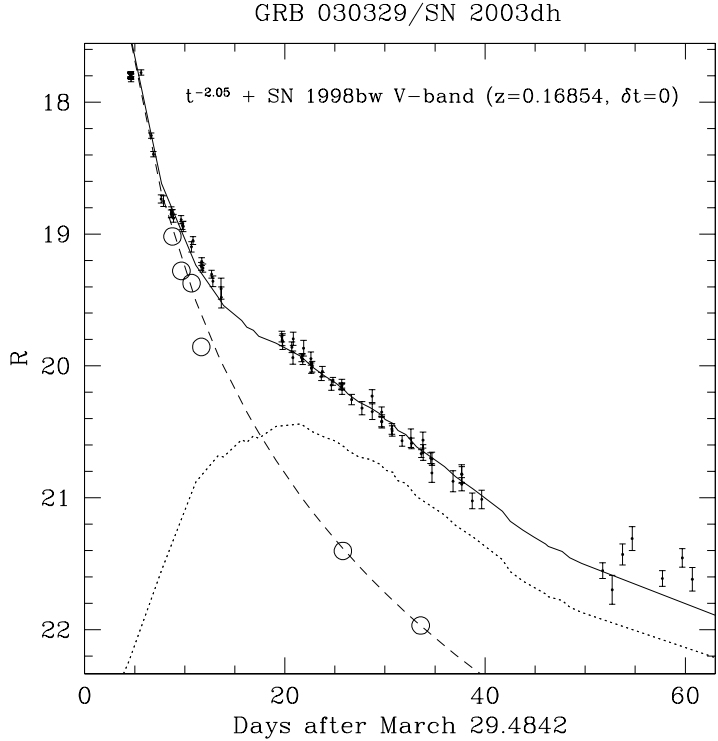}
\end{center}
\caption{Decomposition of the OT $R$-band light curve into the
supernova (\emph{dotted line}) and the power-law continuum
(\emph{dashed line}).  As the light curve model for the supernova, we
took the $V$-band light curve of SN~1998bw (Galama et al. 1998a, b)
stretched by $(1+z)=1.1685$ and shifted in magnitude.  The resulting
supernova light curve peaks at an apparent magnitude of $m_R=20.4$. No
offset in time has been applied between the GRB and the supernova. To
constrain the continuum, information from the spectral decomposition
was used (\emph{big open circles}).
}
\label{fig:decomp}
\end{figure}

The totality of data contained in this paper allows us to attempt to
decompose the light curve of the OT into the supernova and the
afterglow (power-law) component. From the spectral decomposition
procedure described above, we have the fraction of light in the
$BR$-bands for both components at various times, assuming that the
spectrum of the afterglow did not evolve since $\Delta T=5.64$ days.
As we find that the spectral evolution is remarkably close to that of
SN~1998bw, we model the $R$-band supernova component with the $V$-band
light curve of SN~1998bw (Galama et al. 1998a, b) stretched by
$(1+z)=1.1685$ and shifted in magnitude to obtain a good fit. The
afterglow component is fit by using the early points starting at
$\Delta T=5.64$ days with late points obtained via the spectral
decomposition. This can be done in both in the $B$ and in the $R$-band
and leads to consistent results, indicating that our assumption of the
afterglow not evolving in color at later times is indeed valid.

The result of the decomposition of the OT $R$-band light curve into
the supernova and the power-law continuum is shown in Figure
\ref{fig:decomp}.  The overall fit is remarkably good, given the
assumptions (such as using the stretched $V$-band light curve of
SN~1998bw as a proxy for the~SN 2003dh $R$-band light curve).  No time
offset between the supernova and the GRB was applied, and given how
good the fit is, we decided not to explore time offset as an
additional parameter. Introducing such an additional parameter would
most likely result in a somewhat better fit (indeed, we find that to
be the case for $\delta t\approx -2$ days), but this could easily be
an artifact with no physical significance, purely due to small
differences between SN~1998bw and SN~2003dh. At this point the
assumption that the GRB and the SN happened at the same time seems
most natural.

\section{Summary}

The spectroscopy of the optical afterglow of GRB~030329, as first
shown by Stanek et al. (2003a), provided direct evidence that at least
some of the long-burst GRBs are related to core-collapse SNe.  We have
shown with a larger set of data that the SN component is similar
to SN~1998bw, an unusual Type Ic SN.  It is not clear yet whether all
long-burst GRBs arise from SNe.  Catching another GRB at a redshift
this low is unlikely, but large telescopes may be able to discern SNe
in some of the relatively nearby bursts.  With this one example,
though, we now have solid evidence that some GRBs and SNe have the
same progenitors.

\bigskip\noindent 

{\it Acknowledgments} Kris Stanek and Peter
Garnavich were very supportive and equal colleagues in the research
described here.  I would like to thank the many observers who
sacrificed their time to observe this GRB (see Matheson et al. 2003c
for a full list).

\end{document}